\documentclass[final,1p,times]{elsarticle} % do not change
\usepackage{graphicx} % do not change
\usepackage{amssymb} % do not change
\usepackage{amsthm} % do not change
\usepackage{lineno} % do not change

\newcommand{\beq}{\begin{equation}}
\newcommand{\eeq}[1]{\label{#1} \end{equation}}
\newcommand{\ee}{\end{equation}}
\newcommand{\bea}{\begin{eqnarray}}
\newcommand{\eea}{\end{eqnarray}}
\newcommand{\beqar}{\begin{eqnarray}}
\newcommand{\eeqar}[1]{\label{#1}\end{eqnarray}}

\journal{Nuclear Physics A} % do not change
\begin{document} % do not change

\begin{frontmatter} % do not change

%% QM09Author: please enter your  
%% Title, author and address info here; please do not use footnotes

% Your Title - please modify
\title{Light Cone wavefunction approach\\ to open heavy flavor dynamics in the QGP}

% Principle author, and co-authors - please modify
\author{Rishi Sharma$^{a}$ }

% Address - please modify
% note that if you have authors from several institutions, we recommend
% labelling these [a], [b], [c] etc.
\address[a]{Group T-2, Theoretical Division, % label [a]
Los Alamos National Laboratory,
Los Alamos, NM, 87545, USA}

\begin{abstract} % do not change
%% Text of abstract goes here - please modify
%We calculate the lowest order charm and beauty parton distribution and
%fragmentation functions in $D$ and $B$  mesons using the operator definitions of
%factorized perturbative QCD. In the vacuum, we find the leading corrections that
%arise from the structure of the final-state hadrons.
We elucidate the role of time scales that determine heavy quark dynamics in the QGP.
Quark-antiquark potentials extracted from the lattice are used to demonstrate the existence of open
heavy flavor bound-state solutions in the vicinity of the critical temperature, and their light cone
wavefunctions are obtained. We use these wavefunctions to calculate the in-medium  modification of
the heavy quark distribution and decay probabilities.  For the case of high $p_T$ $D$ or $B$ mesons
traversing the QGP, we combine the new meson formation and dissociation
mechanism with the traditional parton-level charm and beauty quark quenching to
obtain predictions for the heavy meson and non-photonic electron suppression in
Cu+Cu and Pb+Pb collisions at RHIC and the LHC, respectively. 
\end{abstract} % do not change

\end{frontmatter} % do not change

%% QM09: we keep linenumbers at least for initial version
%\linenumbers % do not change

%% start of main text - please modify. Below is a sub-set (single section) 
%% of an earlier proceedings that show how one can handle references 
%% and figures etc.
%%\section{}\label{}

\section{Inadequacy of partonic level energy loss for heavy flavor suppression}

It is well known that partonic level energy loss of $c$ and $b$ quarks in the QGP is insufficient to
explain the large suppression $(R_{AA})$ of non-photonic $e^+ + e^-$ measured at
RHIC~\cite{Wicks:2007am}.  To correctly assess the discrepancy between the expected heavy quark
quenching and the measured $R_{AA}$~\cite{Adare:2006nq}, improved treatment of cold
nuclear matter effects~\cite{Vitev:2003xu,Vitev:2006bi} is necessary. It is, therefore,
important  to establish the role of (i) shadowing, (ii) Cronin effect and (iii) cold nuclear matter
energy loss in the analysis of open heavy flavor suppression.  Our results~\cite{SVZ_arxiv} bring
the treatment of cold nuclear matter effects for open heavy flavor production on par with their
implementation in the study of light hadron and direct photon final states~\cite{Vitev:2008vk}.

We now turn to the final-state quark and gluon dynamics. The standard paradigm
for jet quenching is to consider the energy lost by a parton traversing through
the deconfined medium.  This parton then fragments outside the QGP. We perform the energy loss 
calculation in the framework of the GLV approach~\cite{Gyulassy:2003mc}. 
The properties of the soft gluon-dominated matter is constrained by the
experimentally measured charged particle pseudorapidity 
density~\cite{SVZ_arxiv}.

The results for $R_{AA}$ of $\pi^0$ calculated thus, describe the experimental data
beautifully.  On the other hand the experimental results on non-photonic
electrons~\cite{Adare:2006nq}, $R_{AA}^{e^\pm} \sim
R_{AA}^{\pi^0}$ for $p_T >5$~GeV, are in clear contradiction with the small
quenching of $B$ mesons that give an increasingly important ($\geq 50\%$)
contribution to non-photonic $e^+ + e^-$ in this region~\cite{Adil:2006ra,Gang}.
Cold nuclear matter effects increase the discrepancy between the results of
partonic level quenching and experiments~\cite{SVZ_arxiv}.
%In particular, the Cronin effect increases the yield of B mesons for $p_T\sim
%4$GeV and therefore reduces the suppression of non-photonic $e^+ + e^-$.

This discrepancy forces a check of our basic assumptions. Whether or not an energetic quark
traversing the QGP forms a (possibly short-lived) hadronic state in the plasma depends on whether
the formation time of the hadron, $\tau_{\rm{form}}$~\cite{Adil:2006ra}, is less or greater than the
plasma life time.  The small pion mass ensures that the parent light quarks and gluons fragment
outside of the QGP in accord with the traditional picture of jet quenching. The large $D$ and $B$
meson mass, however, implies that charm and beauty quarks will fragment inside the hot 
medium: $ \tau_{\rm form} \propto 1/m_h^2$.  Consequently, the competition between heavy meson
dissociation and the $c$ and $b$ quark decay in the QGP is a likely physics mechanism that may
naturally lead to attenuation of the beauty cross section as large as that for
charm~\cite{Adil:2006ra}.  

\section{Can a $D$ or $B$ meson exist in the QGP?}

\begin{table}[!b]
\begin{tabular}{c|c|c|c|c|c|c|c|c|c}\\
$T$  & $0$  & $0.2T_c$  & $0.4T_c$  & $0.6T_c$  & $0.8T_c$  & $1.0T_c$  & $1.2T_c$  & $1.4T_c$  &
$1.6T_c$\\
\hline
$T$(GeV)   &  $0$ & $0.038$  &  $0.077$  &  $0.115$  &  $0.154$  &  $0.211$  &  $0.230$  &  $0.269$
&  $0.307$ \\
$E_b$(GeV) &  $0.730$ & $0.733$ &  $0.611$ &  $0.256$ &  $0.098$ &  $0.043$ &  $0.031$ &  $0.017$ &
$0.009$\\
$\sqrt{\langle r^2\rangle}$fm & $0.468$ &  $0.466$ &  $0.464$ &  $0.501$ &  $0.632$ &  $0.785$ &  $0.970$ &  $1.263$ &  $1.636$
\end{tabular}
\caption{ Properties of the bound $D$- and $B$-meson bound-state solutions taking the effective light 
quark mass to be $m(T)/\sqrt{2}$. These solutions persist up to temperatures $\approx 1.6 T_c$. }
\label{mqTby2} 
\end{table}

To understand if a $D$ or $B$ meson can exist at rest and in thermal equilibrium
in a QGP, and to understand how the parton distribution functions (PDFs) and
fragmentation functions (FFs) are modified in a thermal medium, we solve the 
Dirac equation for the light quark in a thermally modified Cornell potential. Whether
such thermal effects do indeed take place for the relevant case of an energetic
meson moving through the medium, depends on whether the time taken to equilibrate
this meson is larger or smaller than the time taken to traverse the QGP. We will
address this question in the next section, and for this section consider the
case of a $D$ or $B$ meson in equilibrium with the medium.

Lattice QCD results for the free energy of a system at temperature $T$
containing two infinitely heavy quarks separated by a distance
$r$~\cite{Kaczmarek:2005ui} have been used to extract the static potential
between these quarks. This potential has been employed previously to study the
fate of heavy-heavy bound states in the QGP formed at temperatures above
$T_c$~\cite{Mocsy:2007yj}. We perform a similar study to look at
the possibility of the existence of heavy-light mesons in the QGP by looking for
bound state solutions for the Dirac equation describing the motion of the light
quark in the potential set up by the heavy quark. The temperature up to which
bound states exist depends sensitively upon the details of the potential chosen
and the effective in-medium mass of the light quark. Here we quote results
(Table.~\ref{mqTby2}) for a relatively strongly binding potential and take the
in-medium mass of the light quark to be the $m(T)/\sqrt{2}$.
More details are given elsewhere~\cite{SVZ_arxiv}.  We find that $D$- and
$B$-meson  bound-state solutions persist above $T_c$. Their small binding energy
and large radius, however, will greatly facilitate their subsequent dissolution
in the presence of interactions.

We can use the bound state wavefunctions obtained to calculate the thermally modified PDFs and FFs.
This calculation is most conveniently done using the wavefunctions in the light cone
form~\cite{Brodsky:1997de}. The first results for how the PDFs and FFs change with temperature are
in~\cite{SVZ_arxiv}.
%%%%%%%%%%%%%%%%%%%%%%%%%%%%%%%%%%%%%%%%%%%%%%%%%%%%%%%%%%%%%%%%%%%%%%%%%%%%
\section{Application to heavy meson production in heavy ion collisions}
\label{application}

Heavy flavor dynamics in hot QCD matter critically depends on the time  scales 
involved in the underlying reaction. As discussed 
in the previous section, the dissociation of $D$ and $B$
mesons in the vicinity of $T_c$ can be facilitated by their  small binding 
energy. Whether such thermal effects take place in practice, however, depends 
on the time they need to develop. We can roughly estimate this time by 
boosting  the expanded size of the hadron,  $\approx 2 \sqrt{\langle r^2 \rangle}  $ 
from  Table~\ref{mqTby2}, by the $\gamma$ factor. When compared to  
$\tau_{\rm{form}} $ this time is large and suggests that the fragmentation 
component of the heavy meson dynamics in heavy ion collisions may not be 
affected by the QGP. In what follows we will study this ``instant wavefunction 
limit''.

A meson that is formed and propagates inside the medium will undergo collisional 
broadening and dissociation~\cite{Adil:2006ra,Dominguez:2008be}. Following the diagrammatic
calculation described in~\cite{Adil:2006ra} one can calculate the dissociation time
$\tau_{\rm{diss}}$ for a meson of transverse momentum ${{p_T}}$.

\begin{figure}[!t]
\caption{(Left.) Suppression of $D$ and $B$ hadron production from  meson
dissociation and heavy quark quenching in central Au+Au (top panel)
and Cu+Cu collisions (bottom panel) at $ \sqrt{s_{NN}} =200$~GeV at
RHIC. (Right.) Nuclear modification for the single non-photonic electrons
in central Au+Au and Cu+Cu collisions at RHIC.  Data is from
PHENIX and STAR~\cite{Adare:2006nq}
collaborations. } 
\vspace*{0.4cm}
\includegraphics[width=2.5in,height=2.2in,angle=0]{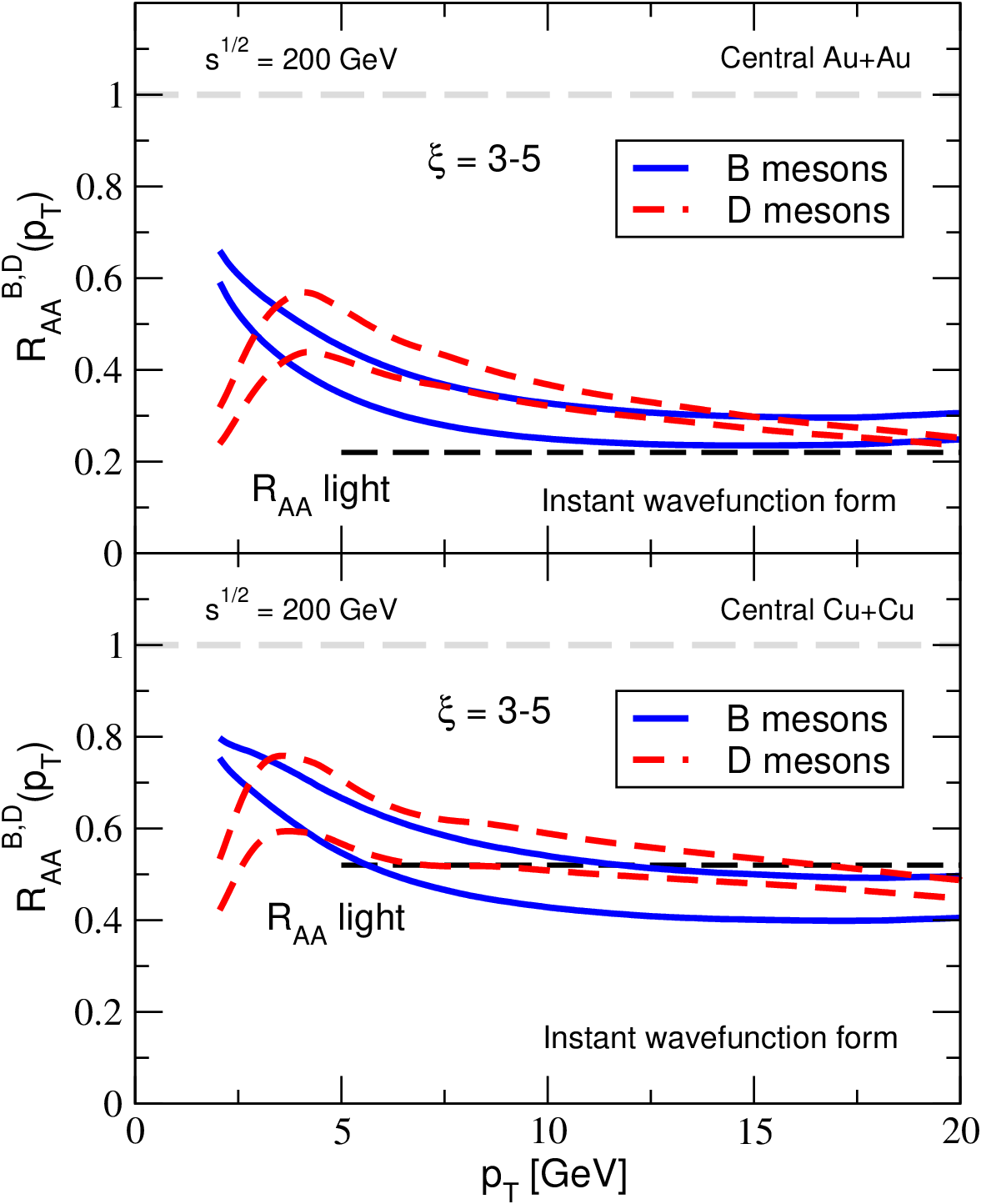}
\includegraphics[width=2.5in,height=2.2in,angle=0]{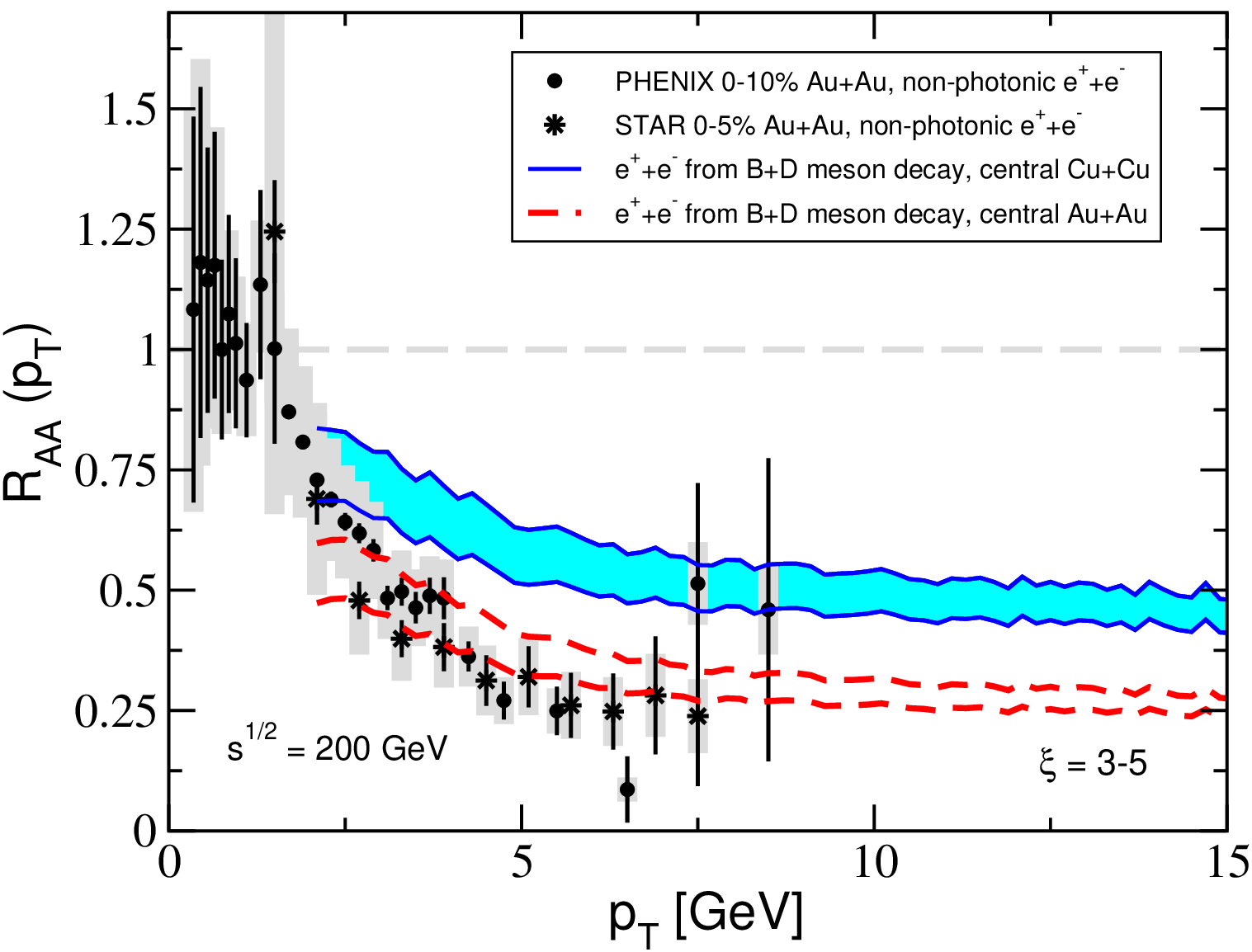}
\vspace*{-0.2cm}
\label{fig:BDSupp-RHIC}
\end{figure}

Employing the rates, $\tau_{\rm{diss}}$ and $\tau_{\rm{form}}$,
the concurrent processes of $c$ and $b$ quark fragmentation and $D$ 
and $B$ meson dissociation are described by % the following set of 
rate equations~\cite{Adil:2006ra} describing the time evolution of quark
and meson ``densities'' ($f^Q(p_T,t)=\frac{d\sigma^Q(t)}{dy d^2p_T}$
and $f^H(p_T,y,t)=\frac{d\sigma^H(t)}{dy d^2p_T}$ respectively).  
At present, there is no reliable way of incorporating the fluctuations
in partonic energy loss in rate or transport equations. Therefore, we include 
the early-time heavy quark inelastic scattering effects approximately as a 
quenched initial condition,
$\;   f^{Q}({p}_{T},0) =
\frac{d \sigma^{Q,{\rm Quench}}}{dy d^2p_T} \;,\;\; 
\;   f^{H}({p}_{T},0) = 0 \; .
$
Here, the attenuated partonic spectrum $\frac{d \sigma^{Q,{\rm Quench}}}{dy d^2p_T}$  
is calculated differentially versus $p_T$ using partonic level quenching. The relevant 
mean quenching time - the time that the  physical system of interest spends in a quark state - can be 
estimated from an analytic solution for $f^{Q}({p}_{T},t), \,
f^{H}({p}_{T},t)$~\cite{SVZ_arxiv}.

\begin{figure}[!t]
\caption{ Suppression of $D$ and $B$ meson production in central Pb+Pb
collisions at $ \sqrt{s_{NN}} =5500$~GeV at the LHC in
two different scenarios. The top panel shows the quenching of heavy hadrons
only due to partonic energy loss. The bottom panel gives $R_{AA}$ for
$D$s  and $B$s with partonic energy loss as well as 
collisional dissociation of heavy meson. Insert shows the corresponding
attenuation of non-photonic electrons in a limited $p_T$ range.} 
\begin{center}
\includegraphics[width=3.0in,height=3.0in,angle=0]{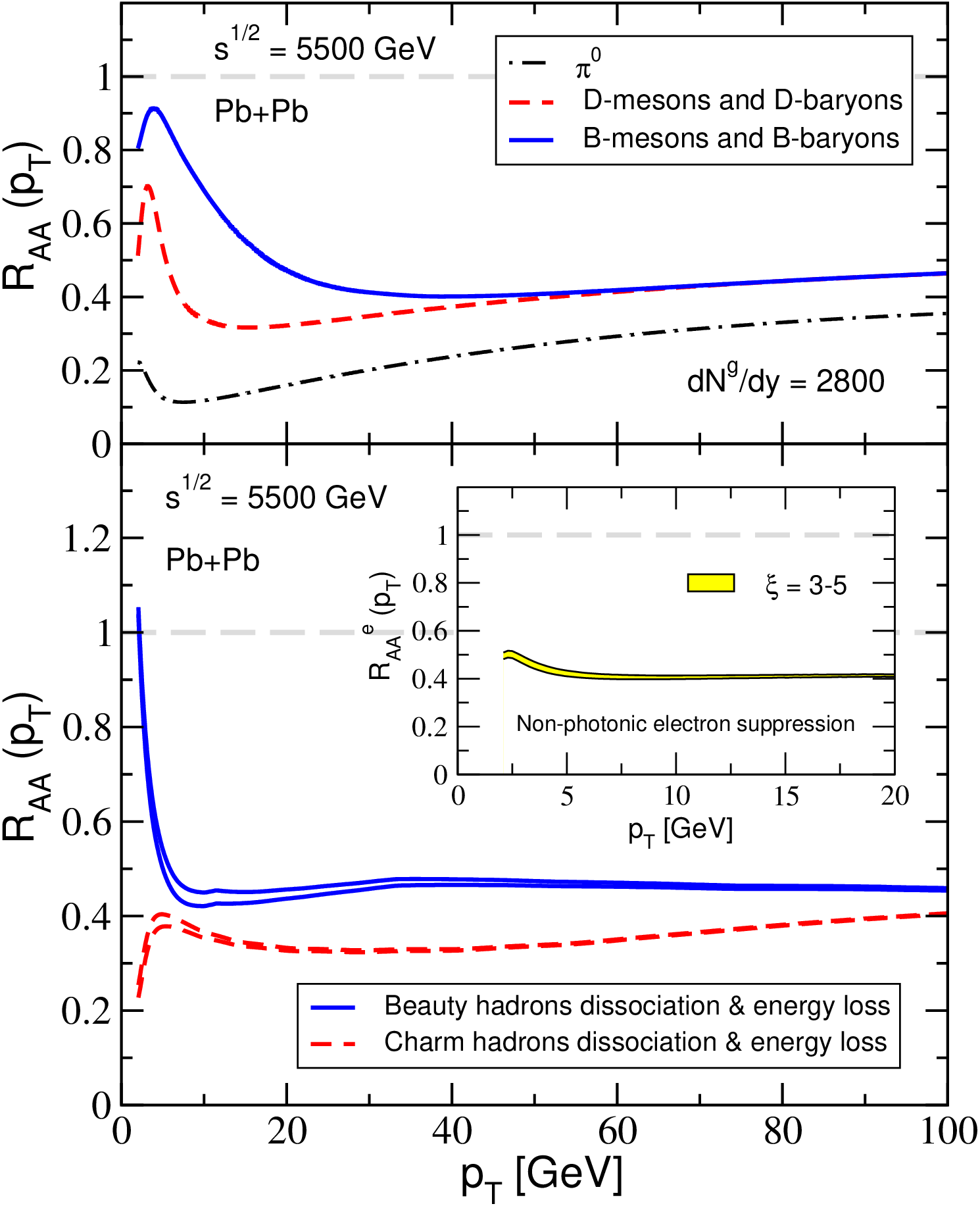}
\end{center}
\vspace*{-1cm}
\label{fig:BDSupp-LHC}
\end{figure}

We integrate numerically the above set of coupled ordinary differential equations 
and  use the same initial soft gluon rapidity  density $dN^g/dy$ as in the 
simulations of $\pi^0$ quenching. The corresponding  suppression of open
heavy flavor in $\sqrt{s_{NN}} =200$~GeV central Au+Au and Cu+Cu collisions 
at RHIC is shown in the top and bottom panels on the left in Fig.~\ref{fig:BDSupp-RHIC}, 
respectively. For $D$ mesons the enhancement from Cronin effect is clearly visible around 
$p_T \sim 4$~GeV.
In both gold and copper reactions at RHIC the suppression  $R_{AA}^B \approx   
R_{AA}^D$  for $p_T > 4$~GeV and these approach the quenching of light hadrons
for $p_T > 10$~GeV. Future vertex detector upgrades at RHIC will 
ensure direct and separate measurements of the $D$ and $B$ mesons and can be 
compared directly with our results. We also give $R_{AA}$ for non-photonic electrons calculated
using semi-leptonic decays of charm and beauty hadrons using
PYTHIA~\cite{Sjostrand:2006za}.
Results for the suppression of open heavy flavor final states  
in central Pb+Pb collisions at $\sqrt{s_{NN}}=5500$~GeV at the LHC taking
$dN^g/dy=2800$ are presented in Fig.~\ref{fig:BDSupp-LHC}.  $R_{AA}$ for
non-photonic electrons is shown in the insert.

%% end of main text

\section*{Acknowledgments} % please check/modify, comment out or delete if not needed
We would like to thank Ivan Vitev and Benwei Zhang for collaboration on the work
and for reading the manuscript. This research is supported by the US Department
of Energy, Office of Science, under Contract No. DE-AC52- 06NA25396 and
by the LDRD program at LANL.
%the NNSF of China and the MOE of China under Project No. IRT0624.
 % do not change 
\end{document}